\documentclass[twoside]{dis08}
\usepackage[latin1]{inputenc}
\usepackage[dvips]{graphicx,epsfig,color}
\usepackage{wrapfig,rotating}
\usepackage{amssymb,amsmath,array}
\def\msb{\overline{\rm MS}}

\begin{document}
\title{Different PDF approximations useful for LO Monte Carlo generators}

\author{A.~Sherstnev$^1$ and R.S. Thorne$^2$
\vspace{.3cm}\\
1 - Cavendish Laboratory, University of Cambridge, 
\\ JJ Thomson Avenue, Cambridge, CB3 0HE, UK\\ 
and Moscow State University, Moscow, Russia (on leave)\\
\vspace{.1cm}\\
2 - Department of Physics and Astronomy, \\
University College London, WC1E 6BT, UK \\
}

\maketitle

\begin{abstract}
The goal of this study is to find a prescription for defining 
parton distributions (PDFs) which are most appropriate for use 
in those codes where only LO 
matrix elements (MEs) are used, as in many Monte Carlo generators.
We describe a modification of LO PDFs, based on using $\alpha_S$ 
at NLO, a specific prescription of the coupling in the QCD evolution, and 
violation of the momentum sum rule. We compare results with  
{\em the truth} -- the prediction using NLO for both MEs and PDFs, and the 
standard LO prediction, finding that the modified PDFs generally produce the 
best results. 
\end{abstract}

\section{Introduction}

The talk \cite{url} considers the use of PDFs in LO Monte Carlo generators. 
It is well known that PDFs extracted at different orders of perturbative 
QCD have large differences in certain regions of $x$. This is because missing 
higher order corrections, both in the parton evolution and in the MEs, play 
an important role in the extraction of the PDFS 
by comparison to experimental data. 
Traditionally, LO PDFs are supposed to be the best choice for use 
with LO MEs as implemented in most Monte Carlo programs. However, another 
viewpoint has recently been put forward, namely it has been suggested that 
NLO PDFs may be more appropriate even for MEs at LO~\cite{Campbell:2006wx}. 
The main justification for this idea is the claim that NLO corrections to 
MEs are small, and the total cross-section changes due to the differences 
in PDFs. 

In this contribution we propose another approach, which combines the 
advantages of both the LO and NLO PDFs. We call the result the LO* PDFs. 
Here we present only two examples of comparisons using all three 
approaches, but  many more examples are available in our previous 
article~\cite{Sherstnev:2007nd}. However, in this article we additionally 
introduce another improvement 
to the modified LO PDFs, namely a change in the scale for the 
coupling used in the evolution of the partons. 
This makes the PDFs more consistent with 
the showering in Monte Carlo codes and has been inspired by feedback 
concerning the original LO* PDF approximation \cite{Seymour}. 


\section{Parton Distributions at Different Orders}
Let us briefly elucidate why the differences between the PDFs at different 
perturbative orders appear. The difference in the gluon PDF is mainly a 
consequence of quark evolution, rather than gluon evolution. The small-$x$ 
gluon is determined by $dF_2/d\ln Q^2$, which is related to the $Q^2$ 
evolution of the quark distributions. The quark-gluon splitting function 
$P_{qg}$ is finite at small $x$ at LO, but has a small-$x$ divergence 
at NLO (and further $\ln(1/x)$ enhancements at higher orders), so the small 
$x$ gluon needs to be much bigger at LO in order to fit data. Differences 
of the same nature appear in quark distributions at LO and NLO. Most 
particularly the quark coefficient functions for structure functions in 
$\msb$ scheme have $\ln(1-x)$ enhancements at higher orders, and the 
high-$x$ quarks are smaller as the order increases. Due to the momentum 
sum rules this is accompanied by a depletion of the quark distribution for 
$x\sim 0.01$. 
This depletion leads to a bad global fit at LO, particularly for HERA 
structure function data, which are very sensitive to quark distributions at 
moderate $x$. In practice the lack of partons at LO is partially compensated 
by a LO extraction of much larger (then at NLO) $\alpha_S(M_Z^2) \sim 0.130$. 
So, the first obvious modification is to use $\alpha_S$ at NLO in a LO fit 
to parton distributions. The problems caused due to the depletion of partons 
have led to a suggestion \cite{Torbjorn} that relaxing the momentum sum 
rule could make LO partons rather more like NLO partons where they are 
normally too small, while allowing the resulting partons still to be bigger 
than NLO where necessary, i.e for the small-$x$ gluon and high-$x$ 
quarks. We call the modification the LO* PDFs. 
The approach does improve the 
quality of the LO global fit. The $\chi^2=3066/2235$ for the standard LO 
fit, and becomes $\chi^2=2691/2235$ for the modified fit with the same data 
set as in~\cite{Martin:2004ir} and using $\alpha_S(M_Z^2)=0.120$ at NLO. 
The momentum carried by input partons goes up to $113\%$. We analysed the 
consequences of these distributions by comparing to cross-sections 
using LO and NLO PDFs in LO Monte Carlo generators and to full NLO results -- 
presenting many examples~\cite{Sherstnev:2007nd}. 

\begin{figure}
\begin{center}
\centerline{
\hspace{-0.8cm}
\epsfxsize=0.48\textwidth\epsfbox{{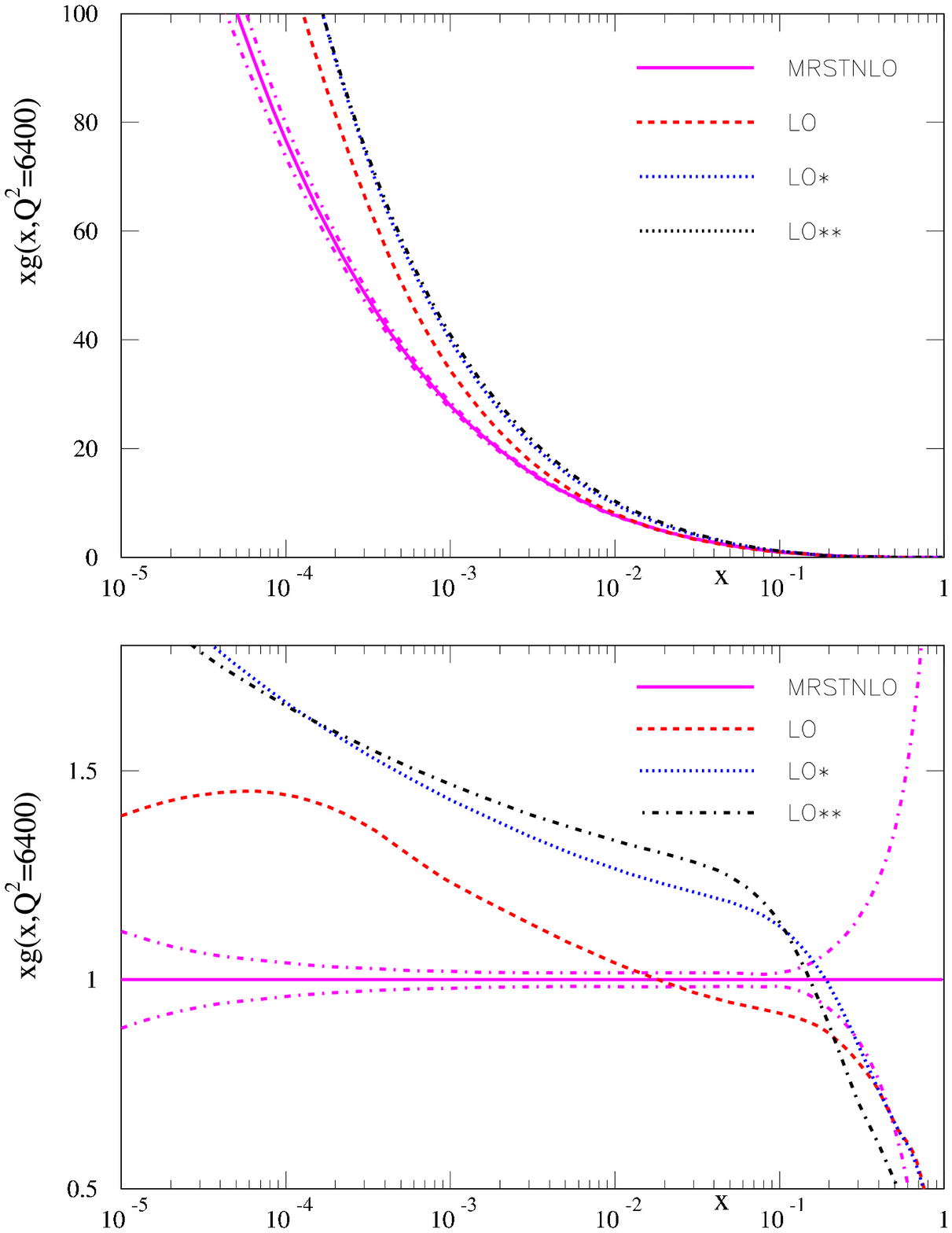}}
\hspace{0.7cm}
\epsfxsize=0.48\textwidth\epsfbox{{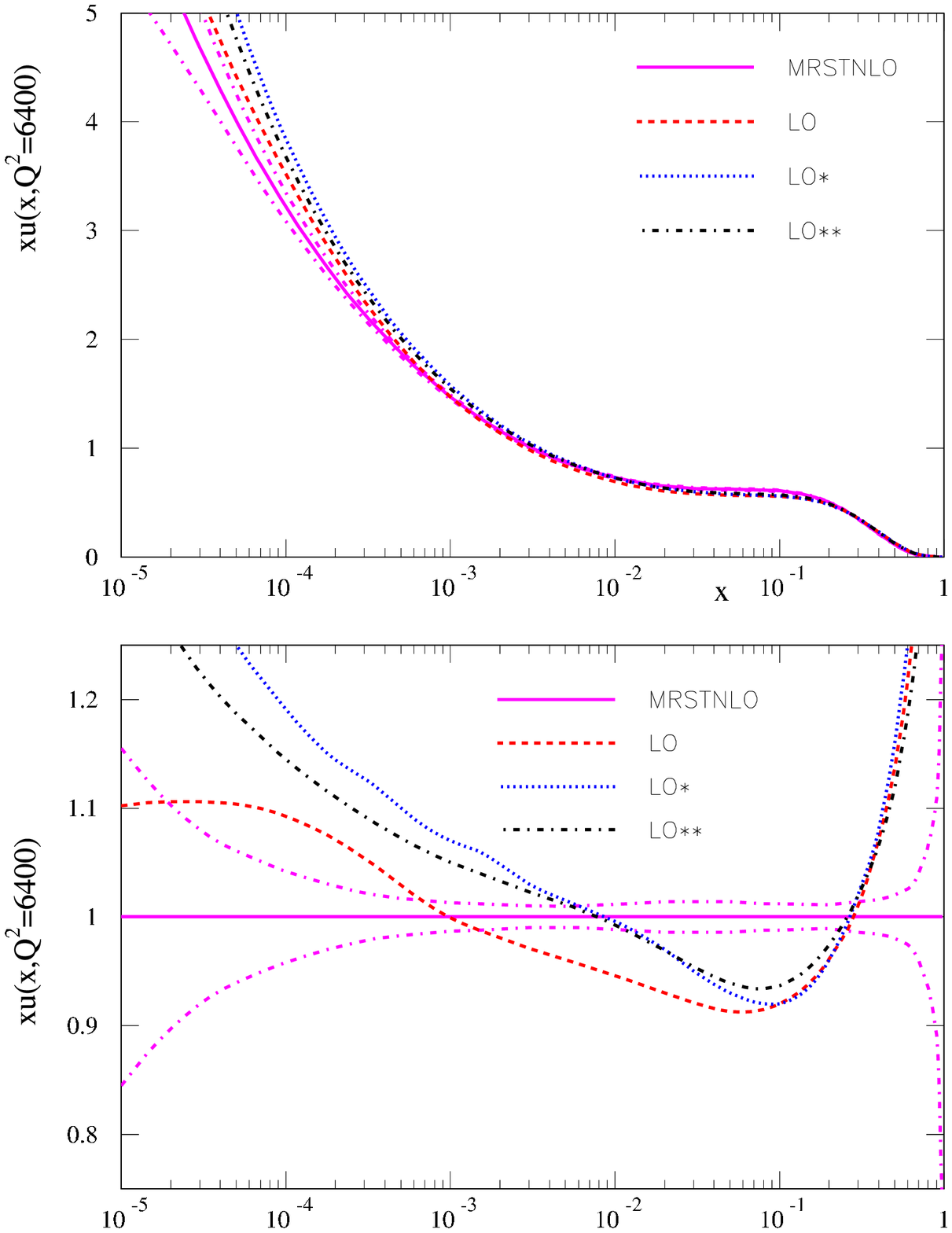}}
}
\vspace{-1.6cm}
\caption{
Parton distribution functions for u-quarks and gluons -- absolute values 
(top) and ratio to NLO (bottom) where the uncertainty band for the NLO PDFs 
is shown. 
}
\vspace{-1.0cm}
\label{pdf}
\end{center}
\end{figure}

Here we introduce an additional modification to the LO* PDFs. 
We replace the standard QCD scale $Q^2$ by $\tilde Q^2=z(1-z)Q^2$
in $\alpha_S$ dependence on the QCD scale in the splitting function $P_{qq}$
(freezing $\alpha_S(\tilde Q^2)$ at 0.5 as $z \to 1$). 
This automatically takes 
into account logarithmically enhanced terms~\cite{Amati:1980ch}. 
The partons obtained using this further 
modification are called the LO** PDFs. As for the LO* PDFS, 
the output is based on the MRST LO PDFs~\cite{Martin:2002dr}. This improves 
the quality of the global fit slightly further, i.e. $\chi^2=2640/2235$,
and the resulting value of $\alpha_S(M_Z^2)=0.115$, smaller than for the LO* 
PDFs. In fig.~\ref{pdf} we illustrate all four PDF 
prescriptions: NLO, LO, LO*, and LO**. LO* and LO** gives more quarks and 
especially gluons for $x<0.1$ than LO. They provide fewer large-$x$ gluons 
than NLO, but more quarks at large $x$ (as for LO) and more quarks and gluons 
than NLO for very small $x$. In general, the LO** modification brings little 
change compared to LO*, mainly a small increase in gluons for 
$x \sim 0.01$, but in many cross-sections this is countered by the smaller 
value of the coupling. 

\begin{wraptable}{l}{0.58\columnwidth}
\centerline{\begin{tabular}{|l|r|r|}
\hline
prescription  & $\sigma_W ({\rm nb})$ & $\sigma_H ({\rm pb})$ \\ \hline
$\rm\sigma(ME_{NLO}\otimes PDF_{NLO})$ & 21.1& 38.0\\\hline
$\rm\sigma(ME_{LO}\otimes PDF_{LO})$ & 17.5  & 22.4    \\
$\rm\sigma(ME_{LO}\otimes PDF_{NLO})$ & 18.6 & 20.3     \\
$\rm\sigma(ME_{LO}\otimes PDF_{LO*})$ & 20.7 &  32.4    \\
$\rm\sigma(ME_{LO}\otimes PDF_{LO**})$ & 20.2  & 35.2    \\
\hline
\end{tabular}}
\caption{Comparison of cross-sections using different prescriptions}
\label{tab:comparison}
\end{wraptable}

\begin{figure}
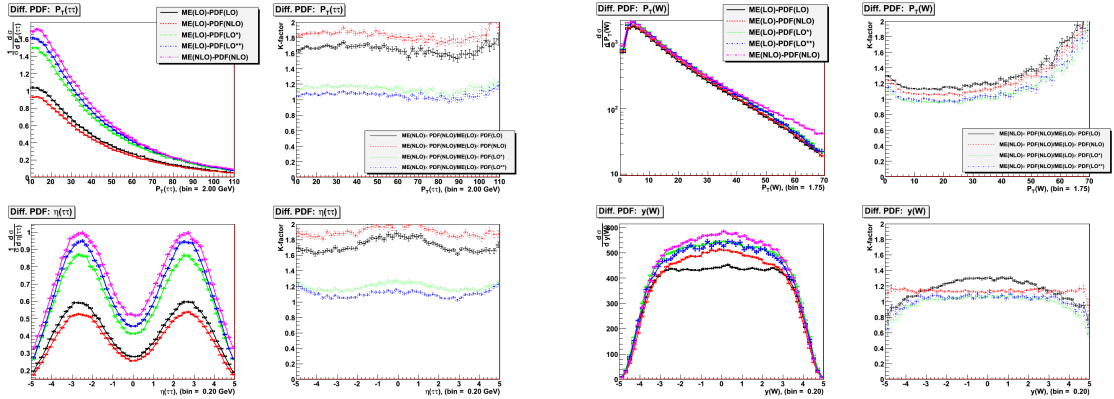

\begin{center}
\centerline{
\hspace{-3.6cm}
\epsfxsize=0.22\textwidth\epsfbox{{c_higgs_ratio.epsi}}
\hspace{4.5cm}
\epsfxsize=0.22\textwidth\epsfbox{{c_wbos_ratio.epsi}}
}
\caption{
Comparison of Z-boson (left plot) and Higgs (right plot) production. {\em The 
truth} and all four approximations are shown. 
}
\vspace{-1.0cm}
\label{examples}
\end{center}
\end{figure}

As an example of the application of the LO** PDFs we consider the  
$W$-boson production 
and the Higgs boson production, both at the LHC. 
For the simulations we used MC@NLO~\cite{Frixione:2002ik}, 
CompHEP~\cite{Boos:2004kh} and FORTRAN HERWIG~\cite{Corcella:2000bw}. 
We compare the use of MEs at LO and the PDFs at LO/NLO/LO*/LO** 
to a combination of PDFs and MEs at NLO (which we call {\em the truth}). 
The total cross-sections for the two examples are shown in 
table~\ref{tab:comparison}. For $W$ production both LO* and LO** are closer to
{\em the truth} than NLO or LO, which is worst, 
but LO** gives a slightly worse answer than LO*.  For Higgs production 
LO** gives the closest number to {\em the truth} and NLO is worst.  
The left of Fig.~\ref{examples} shows two physically important 
distributions for the first example. The LO PDFs show the worst 
behaviour in the $W$-boson pseudo-rapidity, reflecting the PDF depletion for 
moderate $x$. Our LO*/LO** modifications do not have this drawback and 
imitate {\em the truth} much more accurately. The $P_T$ distribution shows 
that we cannot completely simulate the full NLO result with any set of PDFs,
though LO* and LO** give the closest normalizations. 
The right of Fig.~\ref{examples} displays the same distributions for our 
second example. We see that the shapes of all LO approximations are 
fine, but LO** gives a much better normalization.

\section{Conclusions}

We have suggested an optimal set of partons 
for Monte Carlo codes, which is essentially LO but with modifications to make 
results more NLO-compatible. We call the modification LO* and LO** PDFs. 
They are based on three effects: the use of the NLO QCD coupling, relaxing of 
the momentum sum rule, and in the latter case a change in the scale used for 
the argument of $\alpha_S$ for high-$x$ evolution. The resulting PDFs 
are large where it is required for them to compensate for missing higher order 
corrections, but they are not correspondingly depleted elsewhere. 
We have compared in detail the different PDF approximations combined with 
LO MEs to {\em the truth}, i.e. full 
NLO, for two processes which probe different types of PDF, ranges of $x$ and 
QCD scales. One can conclude that, in general, the results are very positive. 
The LO** and LO* PDFs 
provide the best description compared to {\em the truth}. In 
\cite{Sherstnev:2007nd} we saw that this was generally true, especially for 
$s$-channel processes, though LO* gave a slight overestimate for $t$-channel 
processes. We have confirmed that the LO** PDFs give similar results but 
are in most cases even a little closer to {\em the truth} than the LO* PDFs. 
The improvement compared to the LO or NLO PDFs is particularly the 
case in terms of the normalization, 
but the shape is usually at least as good, and sometimes much better, than 
when using LO or NLO PDFs. It should be stressed that no modification of the 
PDFs can hope to reproduce successfully  all the features of genuine NLO 
corrections. In particular we noticed the recurring feature that the 
high-$p_T$ distributions 
are underestimated using the LO generators, and this can only be corrected 
by the inclusion of the emission of a relatively hard additional parton which 
occurs in the NLO matrix element correction. 
However, we propose the use of the LO*, or more correctly, the LO** PDFs 
if only LO MEs are to be used.  Both LO* and LO** PDFs are now 
available in the LHAPDF package~\cite{Whalley:2005nh}, their names are 
MRST2007lomod and MRSTMCal.LHgrid respectively. 

\section{Acknowledgements}
We acknowledge J.~Butterworth, S.~Moch, C.~Gwenlan, P.~Bartalini, 
M.~Cooper-Sarkar, J.~Huston, A.~Martin, S.~Mrenna, T.~Sj\"ostrand, 
J.~Stirling, G.~Watt and B.~Webber for discussions. RST would like to thank 
the Royal Society for the award of a Research Fellowship. AS would like to 
thank the STFC for the award of a Responsive RA position. 

\begin{footnotesize}

\end{footnotesize}

\end{document}